# Hybrid Data-driven Framework for Shale Gas Production Performance Analysis via Game Theory, Machine Learning, and Optimization Approaches

Jin Meng[a,*], Yujie Zhou[a], Tianrui Ye[a], Yitian Xiao[a,**]

[a] *Petroleum Exploration and Production Research Institute, SINOPEC, Beijing, 100083, PR China*

## Highlights

- Shale gas production performance is studied from a data-driven perspective where a novel hybrid analysis framework is proposed.
- Dominant factor analysis, production prediction, and development plan optimization are sequentially performed by integrating game theory, machine learning and derivative-free optimization methods.
- The method is validated with actual production data from the Fuling shale gas field, Sichuan Basin, China and shows good performance in both efficiency and accuracy.

## Abstract

A comprehensive and precise analysis of shale gas production performance is crucial for evaluating resource potential, designing a field development plan, and making investment decisions. However, quantitative analysis can be challenging because production performance is dominated by the complex interaction among a series of geological and engineering factors. In fact, each factor can be viewed as a player who makes cooperative contributions to the production payoff within the constraints of physical laws and models. Inspired by the idea, we propose a hybrid data-driven analysis framework in this study, where the contributions of dominant factors are quantitatively evaluated, the productions are precisely forecasted, and the development optimization suggestions are comprehensively generated. More specifically, game theory and machine learning models are coupled to determine the dominating geological and engineering factors. The Shapley value with definite physical meaning is employed to quantitatively measure the effects of individual factors. A multi-model-fused stacked model is trained for production forecast, which provides the basis for derivative-free optimization algorithms to optimize the development plan. The complete workflow is validated with actual production data collected from the Fuling shale gas field, Sichuan Basin, China. The validation results show that the proposed procedure can draw rigorous conclusions with quantified evidence and thereby provide specific and reliable suggestions for development plan optimization. Comparing with traditional and experience-based approaches, the hybrid data-driven procedure is advanced in terms of both efficiency and accuracy.

# 1. Introduction

Over the last decade, shale gas industry has achieved great economic success and reshaped the world energy market (Middleton et al., 2017; Wang et al., 2014). With the rising demanding of cleaner energy to alleviate climate change, shale gas is expected to play an important role in the low carbon society transition period (Bain & Company, 2021; EIA, 2021). Comprehensive knowledge of the underground situation, advanced drilling and completion technologies, and reasonable production strategies are required to increase profitability of shale gas development (Gregory et al., 2011; Wang et al., 2014; Zou et al., 2017).

Production performance is a vital reference for development plan design and economic potential assessment (Kinnaman, 2011). However, due to the collective effects of complex factors, it is difficult to quantitatively forecast accurate production performance. In recent years, researchers have proposed various methods to analyze shale gas well production, including physical based analytical and semi-analytical methods (Huang et al., 2018; Nobakht and Clarkson, 2012; Yu et al., 2014), reservoir numerical simulation techniques (Cipolla et al., 2010; Shabro et al., 2011; Yu and Sepehrnoori, 2018), decline curve analysis (DCA) approaches (Guo et al., 2016; Tan et al., 2018; Tang et al., 2021), classical data analysis and statistical methods (Morales-German et al., 2012; Xinhua et al., 2020), and data-driven machine learning and deep learning algorithms (Kong et al., 2021; Lee et al., 2019; Mehana et al., 2021; Temizel et al., 2020).

Analytical and semi-analytical solutions for shale gas productions are derived on the basis of fundamental physics laws and fluid flow regimes. Nobakht and Clarkson studied the linear flow behavior in hydraulic fractures under different well control scenarios (Nobakht and Clarkson, 2012). Ogunyomi et al. developed an approximate analytical solution for the dual-porosity model embedding fracture flow and matrix flow, which could be applied across the entire stage of production (Ogunyomi et al., 2015). Huang et al. used a dual porosity model and micro-seismic data to depict fracture network in shale reservoir, and then used Green function and finite difference approach to depict the flow behavior in complex fracture systems (Huang et al., 2018). Yu et al. considered gas desorption effect in the semi-analytical method and tested the method with varying fracture structures (Yu et al., 2014). The flow mechanism in shale gas is very complicated, and includes non-Darcy flow due to inertial effects, transitional flow, slip-flow, diffusion, adsorption, and desorption effects (Qanbari and Clarkson, 2013). Therefore, developing a comprehensive analytical or semi-analytical model is very challenging.

Another widely adopted approach is numerical simulation. The method implies deriving and resolving the governing equations with techniques such as discretization, implicit and explicit formulations, and use of the finite-difference method. Shabro et al. used a pore-scale model that incorporates the effects of no-slip and slip flow, Knudsen

———————

*,** Corresponding author.

*E-mail*: mengjin093@gmail.com (J. Meng *),

      xiaoyitian.syky@sinopec.com (Y. Xiao **)

diffusion, and Langmuir desorption to analyze their individual impact on total production (Shabro et al., 2011). Wu et al. used field measurements to construct an ensemble of complex fracture networks and employed a modified edge-based Green Element Method (eGEM) to speed up the simulation process(Wu et al., 2021). In general, numerical simulation builds the simplified digital mirroring for a physical system by numerically solving a series of governing equations with relevant assumptions. This approach honors the underlying physics fundamentals, but the simulation process may be time-consuming and coarse when the system is highly complex.

Decline curve analysis is one of the most widely used approach to estimate production potential. The empirical decline curve, first introduced by Arps (Arps, 1945), is constructed by fitting production to an empirical equation to estimate the ultimate recovery. Numerous extensions and variants have been proposed since then to improve the applicability and robustness of the method for unconventional resources. Cheng et al. studied the different flow regimes and transient effect of tight and multilayer gas wells and developed a method to decrease the error of traditional DCA (Cheng et al., 2008). Clark et al. used a logistic growth model to predict production for reservoirs with ultra-low permeability and applied volumetric constraints to prevent non-physical forecast (Clark et al., 2011). Duong discovered that for fracture-dominated flow, a log-log plot of rate over cumulative production versus time established a straight-line trend, and therefore can be used to estimate production and ultimate recovery (Duong, 2011). Ogunyomi et al. analyzed the correlation between reservoir, completion parameters, and the key parameters in the empirical model to study the physical basis of DCA models (Ogunyomi et al., 2014). Tang et al. evaluated the performance of seven popular DCA models with field data from Barnett and Marcellus shales, and then proposed a new DCA method, which first transforms the production rate to a log scale and perform nonlinear regression on production data to increase the accuracy of prediction (Tang et al., 2021). In general, DCA is an empirical method which makes predictions based on historical production data. However, the method takes limited consideration of the underlying physical mechanisms.

In recent years, data-driven methods, especially machine-learning-based data analytic approaches are widely applied, precise results are often obtained with high efficiency (Li et al., 2021). Various machine learning algorithms have been successfully applied for production analysis. For instance, random forest (RF), adaptive boosting (AdaBoost), support vector machine (SVM), and neural network (NN) methods perform well for production prediction (Wang and Chen, 2019). The spatial error model (SEM) and the regression-kriging (RK) methods help describing the impact of design choices on well productivity (Montgomery and O'sullivan, 2017) . Gang et al. implemented several tree-based methods, including Gradient Boosting Decision Trees (GBDT) and Extra Tress (ET), to analyze the controlling factors and predict the shale gas production (Gang et al., 2021) . Zhan et al. constructed neural networks for controlling factor evaluation and proved that the method is of great prediction capability (Zhan et al., 2018). Besides supervised learning, Zhou et al. used unsupervised methods, including principal component analysis (PCA) and K-means clustering, to assess the impact of different factors on production performance (Zhou et al., 2014) . As for production optimization, Shelley and Stephenson adopted artificial neural network (ANN) for well-completion optimization to gain more well economic benefits (Shelley and Stephenson, 2000). Nasir et al. optimized well placement by integrating machine learning surrogates with two optimization algorithms, the enhanced success history-based adaptive differential evolution (ESHADE) technique and the mesh adaptive direct search (MADS) technique. The proposed E-MADS method outperforms the two base methods in both accuracy and efficiency (Nasir et al., 2020).

In this study, a novel hybrid data-driven framework for shale gas production performance analysis is developed, where dominant factor analysis, production performance prediction, and development plan optimization are sequentially implemented. Initially, machine learning models and game theory methods are combined to evaluate the dominant geological and engineering factors. Shapley values are then calculated based on the constructed ensemble tree models and the effects on production performance are quantified by the tree-SHAP approximations. Then, a stacked model is trained to predict production, where ensemble tree models are integrated to improve model accuracy and generality. Individual condition expectation (ICE) analysis is carried out on the stacked model to reveal hidden patterns of shale gas development. Finally, based on the data-driven models, the development plan is optimized for maximum productivity by adjusting engineering parameters with the aid of derivate-free optimization methods. The optimization suggestions can provide insight and guidance for designing shale gas development plans. A complete closed-loop workflow embodying analysis, prediction, and optimization is established within the proposed hybrid data-driven framework. The obtained knowledge, models, and suggestions are constantly updated with the enrichment of actual production data.

## 2. Methodology

In this section, key components for the hybrid data-driven framework are introduced in detail, where game theory approaches, machine learning models and optimization algorithms are mutually combined for production performance analysis. At last, an overall workflow integrating all the proposed procedures is elaborated in section 2.4.

### 2.1. Ensemble tree model construction

In this study, geological, engineering and production data are collected and cleaned through preprocessing to construct a high-quality dataset. The goal is to build an accurate and efficient model to predict the shale gas production performance of horizontally fractured wells based upon their geological and development conditions. Various modeling methods can be applied to solve this problem, such as linear regression models, support vector machine regression, decision tree models, and neural networks. Among these methods, tree-based models are relatively interpretable and able to characterize complex, nonlinear relationships between inputs and outputs. A decision tree model entails a series of "if-then" scenarios to repeatedly split the input features and categorizes samples into groups for the purpose of classification or regression. To enhance the performance of a single tree, bagging and boosting strategies are used to generate an ensemble of decision tree models (Dietterich, 2000).

Based on bagging strategies, a collection of models is built in parallel. Training samples are selected randomly with replacement to build individual tree models, so that the models are not related to one another. Bagging can effectively reduce model variance and mitigate overfitting. This study uses the random forest (RF), a classical extension of the bagging method. The random forest technique trains individual tree models with random and independent sampling from training samples and random subset of features (Breiman, 2001). The collection of tree models, with small correlation among one another, is combined to generate the final prediction. In contrast, boosting entails a collection of models built in series, which means that models are trained sequentially with the experience learned from previous models passed on throughout the training period. This study implements two boosting strategies: gradient boosted decision tree (GBDT) and extreme gradient boosting (XGBoost) strategy. For GBDT, a series of weak decision trees are trained at first, and subsequent trees attempt to correct the error from the previous stage by training on the residual error (Friedman et al., 2001). The loss of training is minimized with gradient descent. With this method, GBDT ensures that the training loss is reduced at every step, but the problem of overfitting may arise. XGBoost improved upon GBDT by adding a regularization term to penalize model complexity and randomly selects samples to avoid overfitting (Chen and Guestrin, 2016). In addition, GBDT uses first-order derivative for gradient descent while XGBoost takes the Taylor expansion of the loss function up to the second order to raise the speed and efficiency of gradient descent.

Some important hyperparameters of tree-based models include the number of trees, maximum tree depth, minimum number of samples at each leaf node, subsample size, step size for gradient descent, etc. These hyperparameters have significant influence on model performance in terms of training and testing errors. Inappropriate hyperparameter setting may lead to severe overfitting or underfitting. In this study, random search method is adopted for model hyperparameters searching. Other optimization approaches, e.g., Bayesian optimization as introduced in section 2.4, can also be used, but their applications for hyperparameter search are not the focus of this study.

**2.2. Game theory method for dominant factor analysis**

The knowledge of dominant factors affecting the production performance is vital for shale gas developing and decision making. In past research, dominant factors are usually determined by expert experience, sensitivity analysis through physical modeling, and correlation analysis through production data. Each technique has its own drawbacks: expert experience is usually highly subjective, physical model-based methods are restricted by various ideal assumptions for analytical solutions, and numerical simulations have high computational cost. Crossplots and correlation coefficients (e.g., Pearson, Spearman, Kendall coefficients) are usually adopted for statistical analysis, but they involve little physical information and show poor consistency in shale gas applications due to the ignorance of cooperative effects.

Since ensemble models are constructed to connect production performance with geological and engineering factors, dominant factors can be directly derived by quantifying feature importance of the tree-based models. Global importance values are calculated from an entire dataset, where split count, gain, and permutation are commonly used measures. Split count and gain represent node split frequency and mean decrease impurity correlated to the trees, respectively. They are inconsistent and unreliable measures because only the model structures are evaluated, rather than feature contributions to the outputs. Permutation-based methods quantify global importance by observing the changes in model's error after randomly permuting specific feature columns (Fisher et al., 2019). The methods are consistent but challenging to compute, just like other model agnostic feature attribution methods. In contrast to global importance, individualized importance is calculated for every single shale gas well and provides more insights to physical interpretation. However, apply model agnostic individualized explanation methods to tree-based models are computationally expensive.

In this study, dominant factors are studied from a game theory perspective. Game theory is the study of mathematical models of conflict and cooperation between intelligent rational decision-makers (Myerson, 2013). In a coalitional game, players coordinate their strategies and share the payoff under the rules of a certain game. In shale gas production, each geological and engineering factor can be viewed as a player who makes cooperative contributions to the production payoff within the constraints of physical laws and models. Since the essence of the problems are identical, dominant factors can be very well studied under the game theory framework.

A game $f$ is a function mapping players $x$ to the payoff, while in this study, $f$ is the ensemble tree model, $x_\Lambda$ are shale gas feature data with the set of all collected geological and engineering factors $\Lambda$, and the payoff is the evaluated production performance. In cooperative games, Shapley value is one of the most powerful measures to fairly distribute credit to each individual player (Shapley, 1951). It assigns an importance value for each feature to represent the effect on outputs. For arbitrary feature subset $S \subseteq \Lambda$, the sub-model is retrained on $x_S$ to obtain $f_S$. For a certain feature $i$, the partial effects of withholding it can be quantified by comparing the sub-model differences $f_{S \cup \{i\}}(x_{S \cup \{i\}}) - f_S(x_S)$. Then the Shapley value is defined as the weighted average differences of all possible subsets $S \subseteq \Lambda \setminus \{i\}$:

$$\phi_i = \sum_{S \subseteq \Lambda \setminus \{i\}} \frac{|S|!(M-|S|-1)!}{M!} \left[ f_{S \cup \{i\}}(x_{S \cup \{i\}}) - f_S(x_S) \right], \tag{1.1}$$

where $M = |\Lambda|$ is the number of all features, $\phi_i$ represents the marginal contribution of feature $i$. It can be inferred from Eq. 1.1 that Shapley value is an additive feature

attribution:

$$g(z') = \phi_0 + \sum_{i \subseteq F} \phi_i z_i', \qquad (1.2)$$

where $g$ is the local explanation model, $\phi_0 = f_\varnothing(\varnothing)$, and $z' \in \{0,1\}^M$ represents whether the feature is included in the model. Additivity property indicates that the final output is the summation of the marginal contributions of all features. The solution is unique with three desirable properties: local accuracy, missingness, and consistency (Lundberg and Lee, 2017). These properties guarantee Shapley value a rigorous measurement for feature contribution evaluation. A card game example is discussed in Appendix A to exhibit the standard calculation process of Shapley value.

However, as a model agnostic and individualized feature attribution, computational efforts for Shapley value can be very expensive, especially for high dimensional problems since repetitively model inferencing is required for calculating Eq. 1.1 with exponentially growing complexity of possible feature subset coalitions. To address the problem, Shapley Additivity Explanations (SHAP) value is proposed, which is a convenient approximation of Shapley value and a fast unified measure of feature importance (Lundberg and Lee, 2017). There are several SHAP approximations, where tree-SHAP algorithm is especially designed for ensemble tree models (Lundberg et al., 2020; Lundberg et al., 2018). In this algorithm, SHAP values can be calculated in $O(TLD^2)$ polynomial time, instead of $O(TL2^M)$ exponential time, where $T$, $L$, $D$ are the number of trees, leaves, and the tree depth, respectively. The computational cost is significantly reduced and the tree-SHAP value can be conveniently calculated for practical applications. Readers could refer to (Lundberg et al., 2020; Lundberg et al., 2018) for more details about theoretical proof and algorithm implementation.

In a shale gas application, tree-SHAP value matrix $\Phi_{N \times M} = \{\phi_i^{(n)}\}$, $i = 1, \ldots, M$, $n = 1, \ldots, N$ is calculated, where each element represents the marginal contribution for the $i$ th geological or engineering factor of the $n$ th well observation. The total effect of a certain factor is quantified by its mean absolute contributions over all the $N$ wells:

$$\eta_i = \frac{1}{N} \|\Phi_i\|_1 = \frac{1}{N} \sum_{n=1}^{N} |\phi_i^{(n)}|. \qquad (1.3)$$

$\eta_i$ has the same unit as the output since it quantifies the factor's contribution. Then, dominant factors can be evaluated and sorted by $\eta_i$. In addition, supervised clustering can be performed. In traditional unsupervised clustering, samples are clustered according to various features with very different units or importance, say pressure (MPa) and lateral length (ft). Distance is calculated over all features for comparison although these features are not comparable. With the SHAP values, samples are clustered based on feature contributions, which are physically equivalent and comparable. The supervised clustering can identify groups that share common factors from a revenue perspective and gain more insights to the production process.

Feature contributions can be separated into interaction effects and main effects. How a factor affects the production performance can be studied by examining the separated effects. For arbitrary pairwise features, their Shapley interaction index is defined as (Fujimoto et al., 2006):

$$\varphi_{i,j} = \sum_{S \subseteq \Lambda \setminus \{i,j\}} \frac{|S|!(M-|S|-2)!}{2(M-1)!} \nabla_{i,j}(S), \tag{1.4}$$

$$\nabla_{i,j}(S) = f_x(S \cup \{i,j\}) - f_x(S \cup \{i\}) - f_x(S \cup \{j\}) + f_x(S), \quad i \neq j, \tag{1.5}$$

which represents the impacts of the pairwise features on the model outputs and accounts for most of the variance for Shapley values. It can be naturally extended to approximate the interaction effects with SHAP algorithms introduced before. By freezing each pairwise feature and run tree-SHAP algorithms twice, SHAP interaction value matrix of size $N \times M \times M$ can be obtained in $O(TMLD^2)$ time. Then according to the additivity property (Eq. 1.2), SHAP main effect value of a certain feature is separated from SHAP value:

$$\varphi_{i,i} = \phi_i - \sum_{j \neq i} \varphi_{i,j}. \tag{1.6}$$

There is little dispersion in main effect as it represents the major impact after removing all interaction effects. By studying SHAP main effect value, clear patterns and quantified conclusions for shale gas development can be obtained.

**2.3. Stacked model construction for prediction**

The ensemble tree models embody many excellent attributes. These advantages can be further exploited with the stacking strategy. Stacking, also known as stacked generalization, is one of the most effective ensemble methods applied to improve model performances (Breiman, 1996). The idea of stacking is proposed by Wolpert (Wolpert, 1992), and it has been widely applied to improve prediction accuracy (Chatzimparmpas et al., 2021). In this method, multilevel stacked models are constructed, where predictions from low-level based models are collected and taken as inputs to establish a high-level meta model. Though different aspects are honored in different base models and settings, their advantages and discrepancies can be comprehensively understood and adjusted in the stacked model, and overall model performance is further improved (Pavlyshenko, 2018). The general process of stacked model construction is shown in Fig. 1.

At first, base model architectures $\{f^{(z)}(\boldsymbol{x})\}$, $z = 1, \dots, Z$ are selected, where $f$ can be of various forms, such as linear, tree and ensemble models. The training dataset is then divided into $k$ equal subsets. For a certain base model, $f^{(z)}$ is trained on $k-1$ subsets and validated on the remaining subset iteratively until model predictions of the whole dataset $\xi^{(z)} = f^{(z)}(x)$ are obtained, as shown in the left half of Fig. 1. Then the k-fold cross validation outputs of all base models $(\xi^{(1)},…,\xi^{(Z)})$ are collected as inputs to the high-level stacked model $F$. Although the form of $F$ is not limited, simple linear and tree models are recommended to avoid overfitting. After training the multi-model-fused stacked model, predictions are made on the testing dataset. Since there are $k$ cross validation sub-models of base model $f^{(z)}$, each sub-model is inferenced with all testing data, as shown in the right half of Fig. 1. The average of these $k$ predictions is taken as the integrated feature of $f^{(z)}$ and the procedures are repeated for all base models. Finally, these features are collected and fed into the stacked model to obtain the predictions. In this study, three ensemble tree models are implemented as base models: RF, GBDT, and XGBoost. These ensemble models are powerful and of different characteristics as described in section 2.1. Their advantages are further integrated in a linear meta model to form the high-level stacked model.

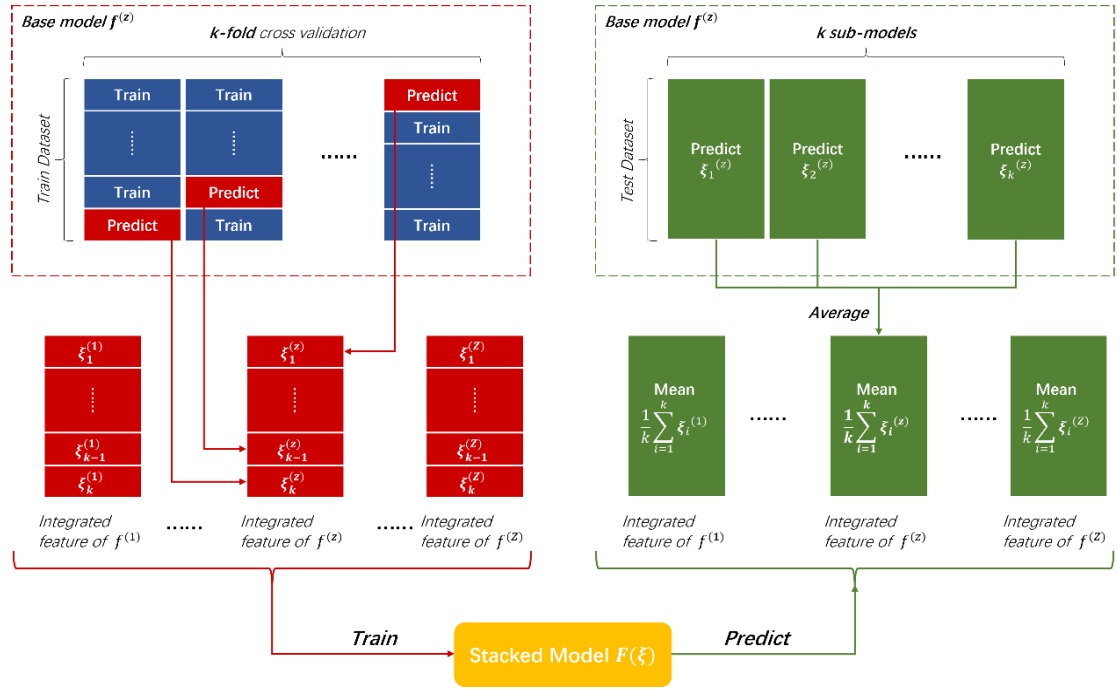

**Fig. 1.** Illustration of stacked model construction and prediction process. Model training process is shown on the left side and testing process on the right side.

The Individual Conditional Expectation (ICE) method can be used to interpret the black box stacked model. As a performance-diagnostic technique, ICE method is developed on the basis of Partial Dependence Plot (PDP) method where local level interpretation is studied (Friedman, 2001). ICE provides insights to global level interpretation of the model by visualization (Goldstein et al., 2015). Correlations between specific features and output predictions are displayed by several equivalent curves, surfaces, volumes, etc. For the feature set $\Lambda$, $S \subseteq \Lambda$ is the subset of features whose impacts on outputs are to be studied, while all the other features $\Lambda \setminus S$ are fixed through the whole process to exclude their interference. Samples $\{(x_S^{(q)}, x_{\Lambda \setminus S})\}$, $q = 1, \ldots, Q$ are then generated, where only $x_S$ are changed within rational ranges in a Cartesian product manner. Production performance can be evaluated for each sample through the stacked model $\hat{y}^{(q)} = F(f, x_S^{(q)})$. With the generated $Q$ data pairs, ICE results are obtained by plotting $\hat{y}^{(q)}$ against the covariate $x_S^{(q)}$ with fixed $x_{\Lambda \setminus S}$ (Goldstein et al., 2015). When $|S| = 1$, for instance, only one geological or engineering factor is studied, ICE visualization is a single curve of $(\hat{y}^{(q)}, x_S^{(q)})$. When two and three factors are studied, surfaces and volumes are inspected, respectively. By interpreting and analyzing the ICE results, hidden patterns of shale gas development are further explored and constructive suggestions for improvement can be proposed. However, for higher dimensionality, i.e., when more than three factors are studied simultaneously, ICE results can hardly be visualized in a brief and comprehensive way.

**2.4. Derivative-free approaches for optimization**

The purpose of optimization is to efficiently and automatically find the optimum development plan that maximizes a designated objective. An optimization problem for shale gas development can be defined as follows:

$$u^* = \arg\max_{u} F(u), \quad s.t. \ u_l \leq u \leq u_u, \tag{1.7}$$

where $u \subseteq x$ is the subset of engineering factors to be optimized such as stimulated length and proppant intensity, $F(u)$ is the objective function to be maximized, $u_l$ and $u_u$ are the lower and upper bounds of $u$. The maximization problem in Eq. 1.7 is usually non-convex and hard to be optimized.

Gradient-based algorithms, such as SGD, Newton's method, and quasi-Newton methods (DFP, BFGS), are the most commonly used optimization methods. In this study, the objective function $F(u)$ is designed as a productivity or economic indicator for the shale gas development. It is computed by the stacked machine learning model in section 2.3, which is a non-linear and non-differentiable black-box function. As a result, gradient-based optimization algorithms that involve computation of the first- and second-order derivatives of the objective function are not applicable. Therefore, three derivative-free global optimization algorithms are integrated in this study to solve the optimization problem in Eq. 1.7. In specific, particle swarm optimization (PSO), differential evolution (DE), and Bayesian optimization (BO) algorithms are applied.

Particle swarm optimization (PSO) is a global stochastic search algorithm introduced by Kennedy and Eberhart (Kennedy and Eberhart, 1995). It is a population-based algorithm that considers a group (swarm) of particles at every iteration, and each particle in the swarm represents a possible solution to the optimization problem. At the $t+1$ iteration step, particle $p$ moves to a new location in the search space:

$$\boldsymbol{u}_p(t+1) = \boldsymbol{u}_p(t) + \boldsymbol{v}_p(t+1), \quad p=1,\ldots,P, \tag{1.8}$$

where $\boldsymbol{u}_p(t+1)$ and $\boldsymbol{u}_p(t)$ are the locations of particle $p$ in the current and previous iteration, $P$ is the total number of particles, and $\boldsymbol{v}_p(t+1)$ is the particle velocity defined by:

$$\boldsymbol{v}_p(t+1) = \omega\, \boldsymbol{v}_p(t) + c_1 \boldsymbol{D}_1\left(\boldsymbol{u}_p^{Pbest}(t) - \boldsymbol{u}_p(t)\right) + c_2 \boldsymbol{D}_2\left(\boldsymbol{u}_p^{Gbest}(t) - \boldsymbol{u}_p(t)\right), \tag{1.9}$$

where $\boldsymbol{v}_p(t)$ is the particle velocity in the previous iteration, $\boldsymbol{u}_k^{Pbest}(t)$ is the personal-best solution of particle $p$ up until current iteration, and $\boldsymbol{u}_p^{Gbest}(t)$ is the global-best solution of all particles. $\omega$, $c_1$, and $c_2$ are the inertial, cognitive, and social hyperparameters, which determine the weights of the three velocity components. The stochastic nature of PSO is preserved in the diagonal matrices $\boldsymbol{D}_1$ and $\boldsymbol{D}_2$, as their diagonal elements are standard uniform random variables and regenerated in each iteration.

Differential evolution (DE), similar to PSO, is a population-based global stochastic search algorithm introduced by Storn and Price (Storn and Price, 1997). A population consists of $P$ parameter vectors, and each vector represents a feasible solution. The parameter vectors are updated in each iteration through the mutation, crossover, and selection steps. In mutation step, three vectors are randomly selected to generate a mutant vector:

$$\boldsymbol{w}_p(t+1) = \boldsymbol{u}_{p_1}(t) + J \cdot (\boldsymbol{u}_{p_2}(t) - \boldsymbol{u}_{p_3}(t)), \tag{1.10}$$

where $J \in [0,2]$ is a constant factor which controls the amplification of the differential variation. In crossover step, the trial vector is generated to increase the diversity:

$$\boldsymbol{v}_p(t+1) = \left(v_{1,p}(t+1),\ldots,v_{D,p}(t+1)\right) \tag{1.11}$$

$$v_{d,p}(t+1) = \begin{cases} w_{d,p}(t+1) & \text{if } rand(d) \leq CR \\ u_{d,p}(t) & \text{if } rand(d) > CR \end{cases}, \quad d=1,\ldots,D, \tag{1.12}$$

where $rand(d)$ is a stand uniform random variable, and $CR \in [0,1]$ represents the crossover probability. In the selection step, the trial vector $v_p(t+1)$ is compared with current solution $u_p(t)$ using the greedy criterion. The parameters are updated only when the original design is improved:

$$u_p(t+1) = \begin{cases} v_p(t+1), & \text{if } F(v_p(t+1)) > F(u_p(t)) \\ u_p(t), & \text{otherwise} \end{cases}. \quad (1.13)$$

Bayesian optimization (BO) is a highly efficient derivative-free and machine learning based technique to locate global optimal solution according to Bayesian posterior probability. Direct derivation of posterior probability from an objective function is impractical, thus a surrogate is utilized instead. The surrogate is usually constructed by a Gaussian process (GP), whose power to express a rich distribution relies on the covariance function. An acquisition function is defined to find the next potential optimal solution out of the posterior probability. The solution is then added to the training sample and the surrogate is retrained for next iteration until specific convergence conditions are reached. The BO method is a quite flexible framework for optimization. The expressiveness of a GP for different distributions can be further extended by introducing various covariance functions, such as squared exponential kernels and Matérn class functions. Besides GP, various machine learning models can also be applied for surrogate construction. There are many alternative definitions for acquisition function, such as probability improvement, expected improvement and entropy search. The BO method can be customized and modified according to actual demand. More details and modifications can be found in (Shahriari et al., 2015; Snoek et al., 2012).

## 2.5. Hybrid data-driven framework

In this study, a novel framework for shale gas production performance analysis is proposed. In this hybrid data-driven workflow, dominant factor analysis, production prediction and development optimization are sequentially performed, where the machine learning methods, game theory approaches and optimization algorithms introduced above are integrated in all procedures. The flowchart of the closed-loop workflow is shown in Fig. 2.

At first, geological factors including reservoir properties, engineering factors including drilling and completion data, and production performance characterization data such as test production, cumulative production, and estimated ultimate recovery (EUR) are collected. The data are cleaned via preprocessing to form a high-quality dataset, which is the foundation of the whole study. Then ensemble tree models are trained on the prepared dataset, where RF, GBDT, and XGBoost models are constructed in this study. We calculate Shapley values for these three base models using the tree-SHAP algorithm. With the Shapley values, dominant factors are sorted by their contributions and the development patterns for a specific factor or a production well can be studied. The multi-model-fused stacked model is then constructed from the three base models for production prediction. With the well-trained stacked model, ICE curves, surfaces, and volumes are plotted to study the effects of single, double, and triple factors on the outputs. At last, derivative-free optimization methods are applied for development plan optimization, where PSO, DE and BO algorithms are compared in this study. The analysis, prediction, and optimization procedures make up a closed loop for shale gas development study and the obtained results are iteratively validated and updated with the enrichment of the collected field data.

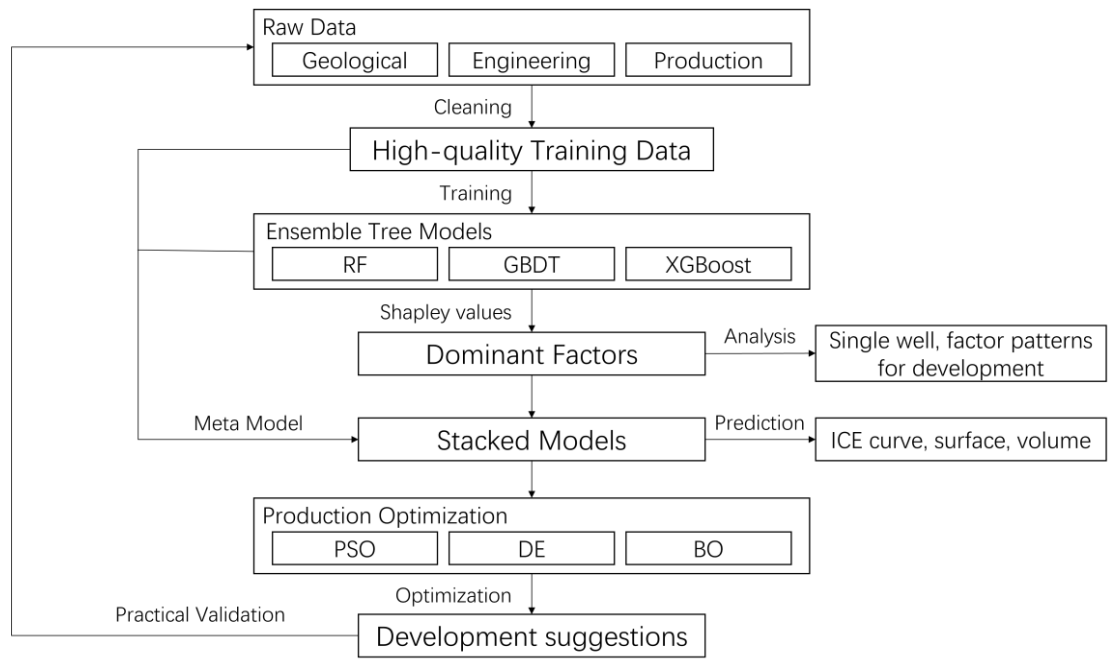

**Fig. 2.** Flowchart of hybrid data-driven framework workflow.

## 3. Results and Discussions

In this section, we validate the hybrid workflow proposed in section 2 with field data collected from the Fuling shale gas field. Backgrounds of the studied area and basic information of the dataset are described in section 3.1. Dominant factors and development patterns are analyzed in section 3.2. Stacked model is constructed in section 3.3 to predict production and plot ICE visualizations. At last, development suggestions are provided by optimizing engineering factors in section 3.4.

### 3.1. Dataset preparation

The Fuling shale gas field is the first major commercial shale gas project in China and the largest shale gas field outside North America. The field was discovered in 2012 and put into development in 2013. The daily gas production from the field exceeded 20 million cubic meters in 2020. Gas production mainly comes from the high-quality marine shale in the upper Ordovician Wufeng formation and lower Silurian Longmaxi formation. This study focuses on the 254 shale gas wells from the first period production project of the Jiaoshiba area located in the southern Fuling district. Fig. 3 shows a map overview of the Fuling field and Jiaoshiba area. The Jiaoshiba main area can be further divided into four sectors according to tectonic structure, fracture distribution, and resource concentration. The main sector occupies the largest area and its average production is the highest, while southwest sector is the lowest. The production performance varies a lot in different sectors, which is mainly caused by differing geologic conditions. The data collected from the field include the geological, drilling, and completion parameters, which are summarized in the Appendix B.

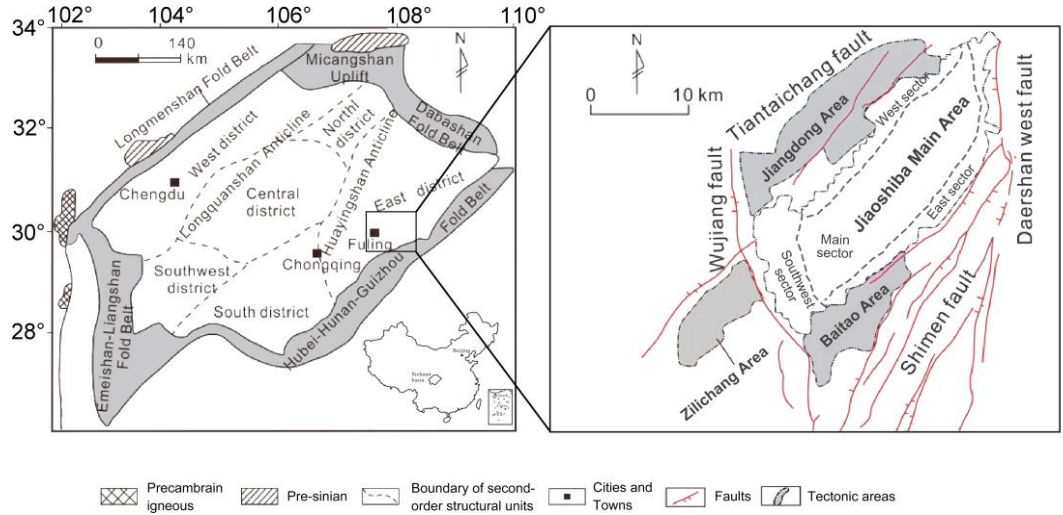

**Fig. 3.** Schematic location of the Fuling shale gas field in Sichuan Basin and tectonic units of Jiaoshiba area (modified after (Yang et al., 2019) and (Xu et al., 2020)).

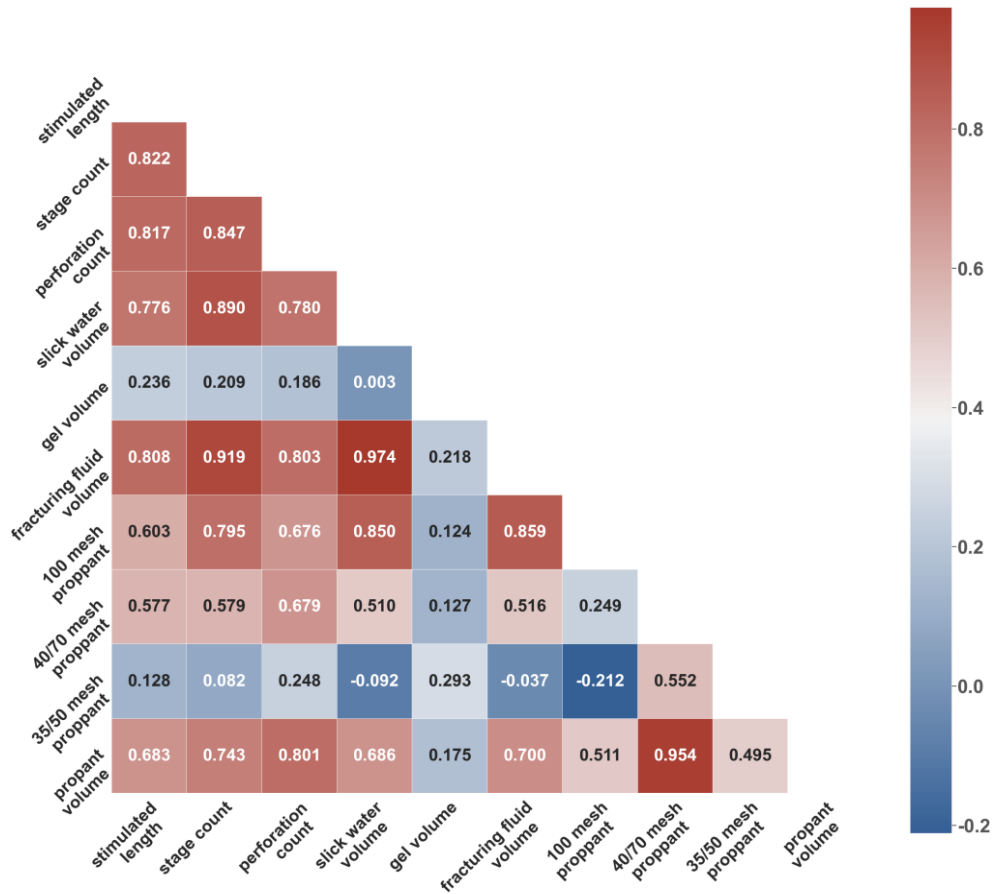

**Fig. 4.** Correlation heatmap of completion and fracturing factors.

Data preprocessing is the essential step to obtain high quality dataset, which is the foundation of the following analysis. Since the sample size is relatively small (254 samples in total), the maximum number of input features should be limited to prevent overfitting. The application of data preprocessing is problem-oriented and customized for different datasets. In this study, features with many missing values are discarded. Common data completion and interpolation methods are not applied because they may introduce noise and uncertainty for model constructions due to the lack of physical control. Samples with abnormal values that deviate a lot from the main body are discarded from the dataset. Finally redundant features are removed in consideration of expert suggestions and statistical analysis, and the Pearson correlation coefficients are computed for all features. A correlation heatmap for completion and fracturing factors are shown in Fig. 4 for illustration. For instance, the 40/70 mesh sand volume is highly correlated to the total proppant volume with a correlation coefficient over 0.90. This is because 40/70 mesh sand is the major component of the proppant used in the study area. The fracturing fluid volume is highly correlated to stimulated length, stage and perforation counts due to the nature of hydraulic fracturing. These redundant features are discarded or integrated to generate meaningful interaction features based on expert knowledge, such as fracturing fluid volume over stimulated length representing the fluid intensity.

**Table 1.** Summary of input features and target of the dataset after preprocessing.

| Category | Name | Unit | Optimizable | Notes |
|---|---|---|---|---|
| Geologic | formation depth | $m$ | N | |
| | TOC | % | N | |
| | porosity | % | N | |
| | hydrocarbon saturation | % | N | |
| | tectonic curvature | - | N | representation of natural fracture development |
| | formation pressure coefficient | - | N | formation pressure over hydrostatic pressure |
| | breakdown pressure | $MPa$ | N | |
| Drilling | target layer penetration | % | Y | |
| | angel to minimum horizontal stress | ° | Y | |
| Completion | stimulated length | $m$ | Y | |
| | stage count | - | Y | |
| | fracturing fluid intensity | $m^3/m$ | Y | fracturing fluid volume over stimulated length |
| | proppant intensity | $m^3/m$ | Y | proppant volume over stimulated length |
| Production | EUR | $10^8 m^3$ | - | evaluated from long term production history |

The estimated ultimate recovery (EUR) is chosen as the learning target in this study, since it is an accurate long term quantitative indicator for production performance of shale gas. The final processed dataset contains 245 samples, each sample consists of 13 geological and engineering factors correlated to the EUR, are summarized in Table 1.

**3.2. Dominant factor analysis**

Machine learning models are initially trained to study the dominant factors of EUR. Three ensemble tree models, random forest (RF), gradient boosted decision tree (GBDT) and extreme gradient boosting (XGBoost) models are constructed and compared. Models are trained on 80% of the data and the remaining 20% are left for validation. Hyperparameters are tuned by random search with 5-fold cross validation to reach higher accuracy and avoid overfitting. The training and validation accuracy of these ensemble models are summarized in Table 3, which will be further discussed in detail in section 3.3.

At first, we study the feature importance with traditional correlation analysis. For illustration, Pearson and maximal information coefficient (MIC), which are typical measurements for linear and non-linear correlations respectively, are calculated for each factor. The grey relational analysis (GRA) coefficient, which measures the similarities of sequences, is also computed for comparison purposes. The factors are ranked according to their correlations, representing their importance to the EUR. As shown in Table 2. the rankings of the three methods vary a lot with each other. In fact, such inconsistency can be frequently observed when using different correlation-based methods to evaluate the relationships of two variables, especially when they are not linearly correlated. Moreover, mutuality effects of these factors can't be reflected in the simple correlation coefficients. As a result, the conclusions are usually inconsistent and misleading.

We then study the dominant factors with the aid of the game theory method. Based on the trained machine learning models, Shapley values for each factor in each model are calculated by the SHAP algorithm (Fig. 5). Since the contribution to EUR is quantified by the SHAP value, each row represents the impact of a certain factor. The color represents the numerical value of the factor, with red indicating high and blue indicating low. For example, according to the RF model (Fig. 5a), formation depth may cause at most $-0.7\sim0.4\times10^8 m^3$ changes of EUR. All the factors are ranked from top to bottom according to their influences. Comparing Fig. 5a, 5b and 5c, although the models are very much different from each other, the SHAP value distribution and tendency of each factor are quite similar, showing that the game theory methods are robust and consistent in this application.

**Table 2.** Factor rankings of feature importance evaluated by different methods.

|  | Pearson | MIC | GRA | SHAP analysis | | |
| --- | --- | --- | --- | --- | --- | --- |
|  |  |  |  | RF | GBDT | XGBoost |
| formation depth | 1 | 1 | 9 | 1 | 1 | 1 |
| hydrocarbon sat. | 5 | 2 | 1 | 2 | 2 | 2 |
| pressure coeff. | 2 | 3 | 8 | 3 | 4 | 3 |
| breakdown pressure | 4 | 6 | 12 | 4 | 3 | 4 |
| porosity | 3 | 4 | 11 | 5 | 5 | 5 |
| stimulated length | 10 | 5 | 6 | 6 | 6 | 6 |
| angle to Hmin | 11 | 12 | 10 | 7 | 7 | 7 |
| fluid intensity | 9 | 9 | 4 | 8 | 8 | 12 |
| TOC | 7 | 8 | 7 | 9 | 9 | 10 |
| target layer pen. | 8 | 10 | 13 | 10 | 12 | 11 |
| curvature | 6 | 7 | 5 | 11 | 10 | 9 |
| stage count | 12 | 13 | 2 | 12 | 13 | 13 |
| proppant intensity | 13 | 11 | 3 | 13 | 11 | 8 |

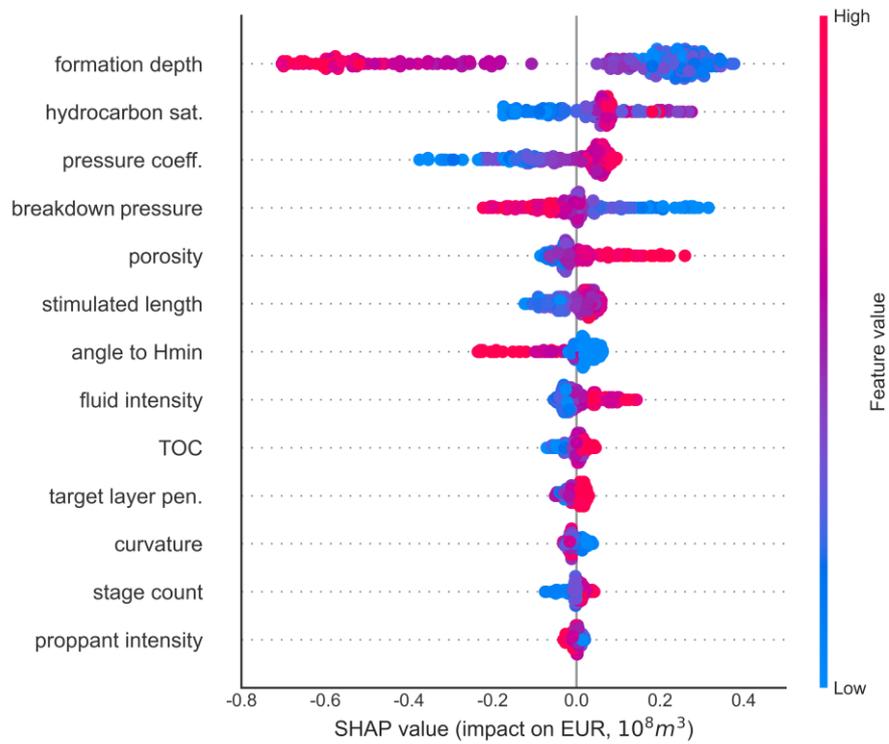

(a)

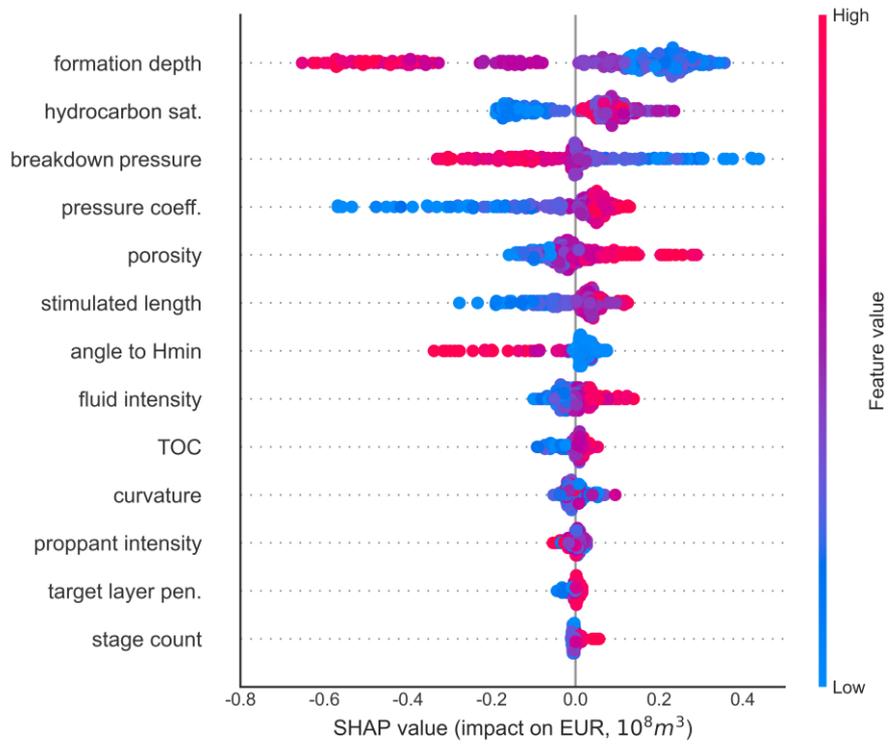

(b)

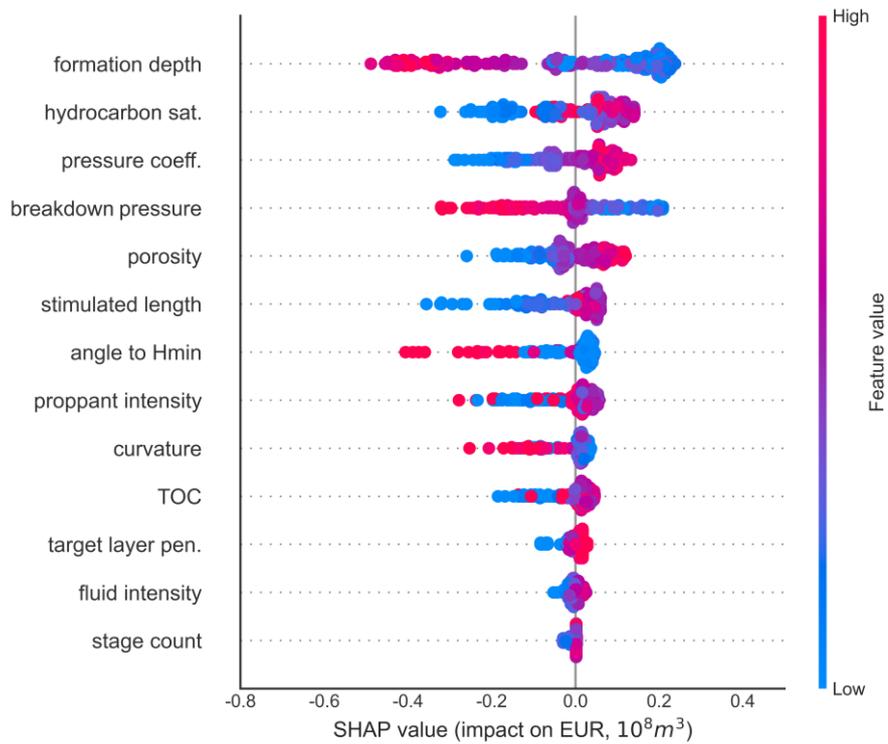

(c)

**Fig. 5.** SHAP value summary plot for different geological and engineering factors obtained from (a) RF, (b) GBDT and (c) XGBoost models. Each dot represents the SHAP value of a certain factor from a certain well sample.

We then calculate the mean absolute SHAP values (Eq. 1.2) to represent the total impact of a factor, from which the dominant factors can be ranked (Table 2). The dominant factors and their rankings in different models are almost the same, suggesting that geological attributions are the most important factors affecting the shale productivity in this study area. Comparing to traditional correlation coefficients, the SHAP analysis method shows great consistency since it measures the contributions of each factor rather than mere correlations. Therefore, the results are much more convincing and physically interpretable. The mean absolute SHAP values of the first six dominant factors are shown in Fig. 6. Around 30% contribution to the EUR comes from the formation depth, which is the most influential factor in this study area. Stimulated length is the most significant engineering factor. These factors contribute over 80% to the EUR in total, and their rankings are identical in each model. As for other factors, the results are slightly different due to the diversity and fluctuation of different models. However, since the major components have been captured by SHAP analysis, the slight fluctuations won't disturb the consistency and effectiveness of the conclusions.

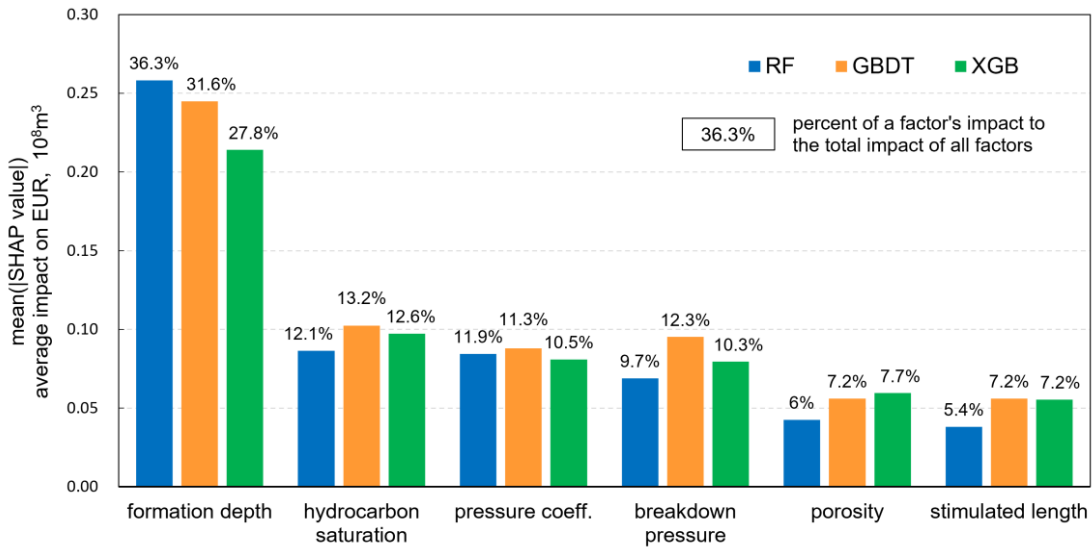

**Fig. 6.** Mean absolute SHAP values for the first six dominant factors of three ensemble models.

The EUR impact of the changing pattern for any specific factor can be further investigated. For example, crossplots of EUR versus one-to-one relationships of four dominant factors are shown in Fig. 7. Since EUR is cooperatively affected by multiple factors, we struggle to find clear patterns or draw quantified conclusions from these crossplots. Then we calculate the SHAP main effect values (Eq. 1.4~1.6) for each factor to reduce interferences. According to the SHAP additivity (Eq. 1.2), actual contributions of each factor are fluctuated around the EUR baseline $\phi_0 = f_\varnothing(\varnothing)$. For brevity, the baseline value which is equal to mean EUR is removed in the SHAP dependency plots (Fig. 7). The effects of these factors are clearly displayed and instructive conclusions can be obtained. As indicated in Fig. 7b, formation less than 2,750$m$ deep may cause $0.3 \times 10^8 m^3$ elevation of EUR over the baseline. But with the formation going deeper, the positive effects declined immediately and turned to negative effects. Formation over 3,000$m$ depth may cause $0.6 \times 10^8 m^3$ EUR drop down from the baseline. Shallow formation under 2,800$m$ is probably the most favorable zone for shale gas accumulation and production. Hydrocarbon saturation represents resource concentration which has significant positive effects on EUR (Fig. 7d). But a plateau is reached after 20%, indicating that other conditions need to be considered to further promote the resource recovery rate. Pressure coefficient which influences the hydraulic fracturing also exhibits similar phenomenon (Fig. 7f). Stimulated length is the most significant completion factor in this study area. Although numerical simulation often suggests the longer stimulate length the better, data analysis from the field shows that there is not much growth potential by simply pursuing super long horizontal wells (Fig. 7h), probably due to the higher construction costs and failure rate.

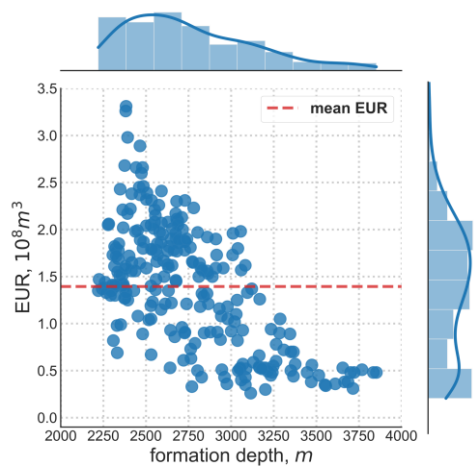

(a)

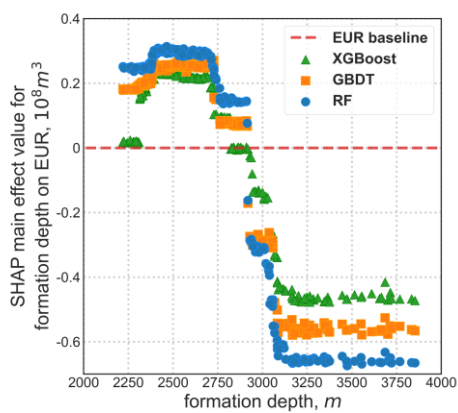

(b)

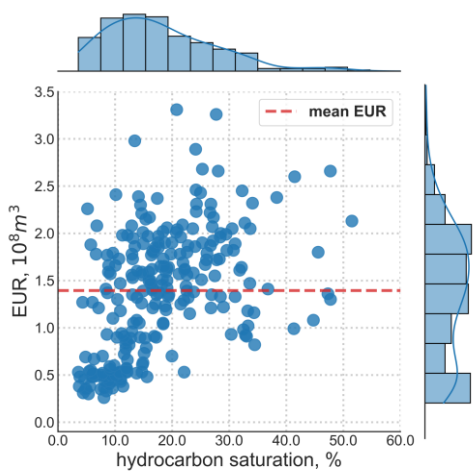

(c)

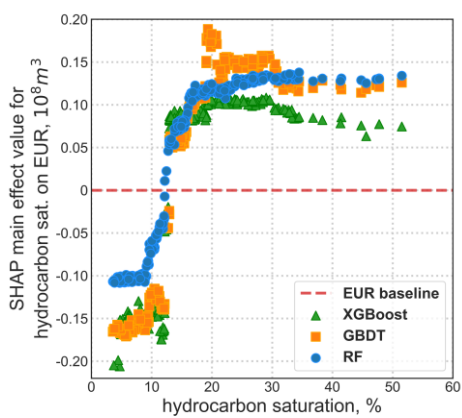

(d)

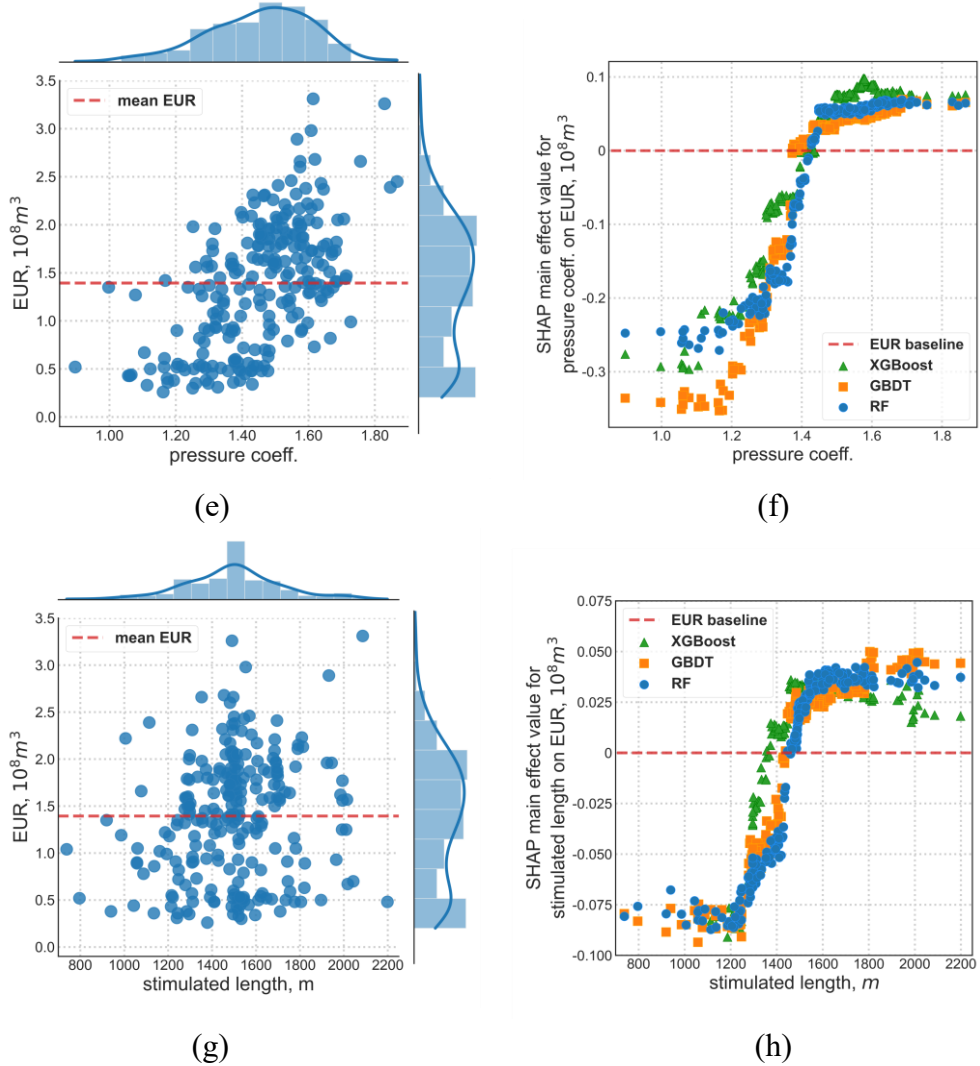

(e)  (f)

(g)  (h)

**Fig. 7.** Comparisons of traditional crossplots and the SHAP dependency plots of four dominant factors: (a)(b) formation depth, (c)(d) hydrocarbon saturation, (e)(f) pressure coefficient and (g)(h) stimulated length.

Dominant factors can also be studied for each production well. Since the EUR output can be decomposed into the base value (equal to mean EUR) and the summation of all SHAP values (Eq. 1.2), favorable and unfavorable factors can be directly distinguished and quantified according to their contributions. Three representative wells are shown in Fig. 8 for comparison. In Fig. 8a, the well contains various advantages, such as shallow formation depth, low breakdown pressure, large porosity and hydrocarbon saturation. These favorable factors push the EUR of this well much higher than the regional base value. The production may be further improved if the stimulated length can be longer since it is the major source of negative effects. Contributions of favorable and unfavorable factors are evenly matched in Fig. 8b, thus this well is close to the average. Fig. 8c shows a very deep well with poor geological conditions. The unfavorable factors greatly pull down the EUR. With our newly-devised approach and capability, one can efficiently evaluate the status of every production well, quantitatively interpret what factors cause the production differences, and gain deeper understanding for the exploration and development.

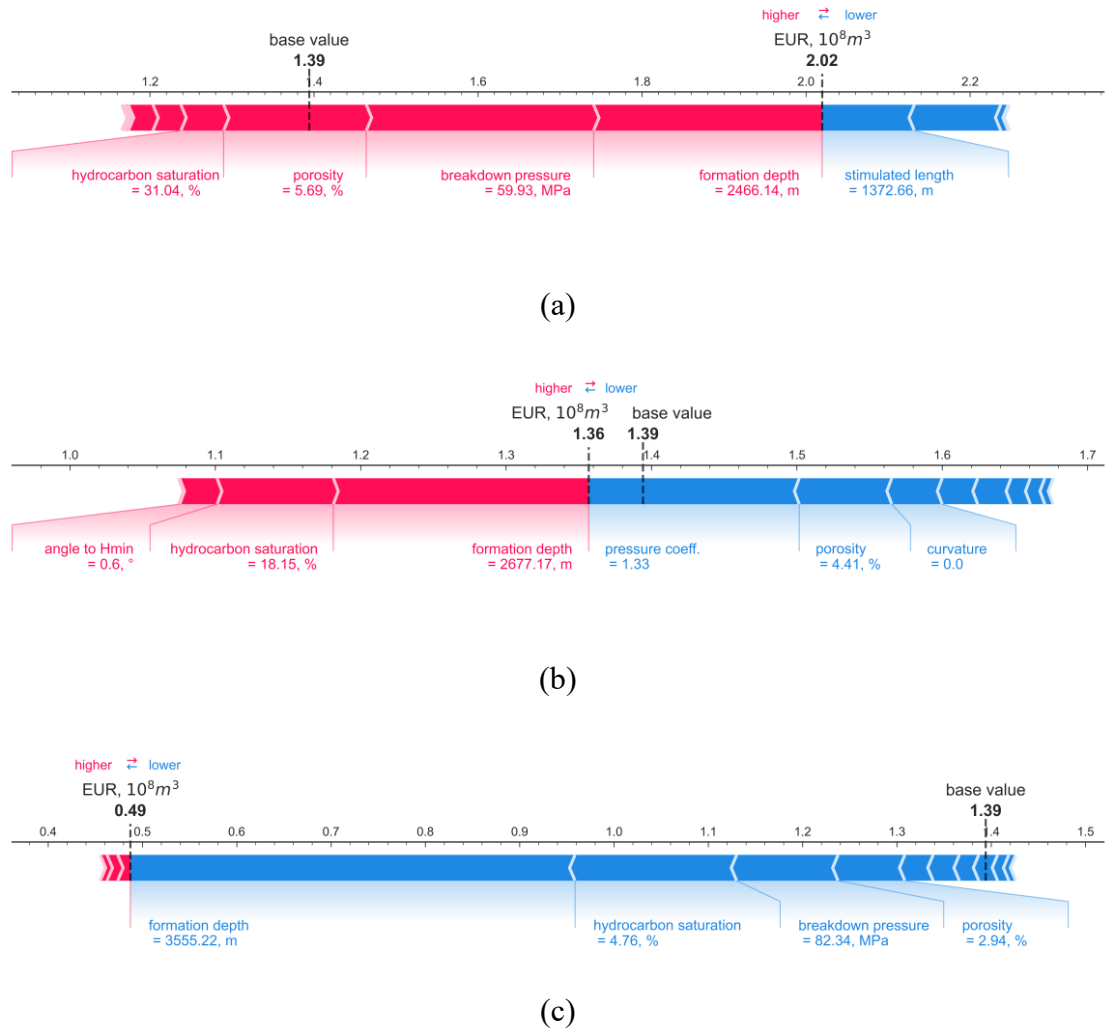

(a)

(b)

(c)

**Fig. 8.** SHAP explanations for the EUR of three shale gas wells in the study area. Red bar represents positive effects on EUR and its length represents the amount of contribution, while blue represents negative.

### 3.3. Production performance prediction

In this section, a stacked model is trained to predict EUR based on the three machine learning models trained in section 3.2. At first, the performances of the three base models are compared. As shown in Fig. 9a, 9b and 9c, the prediction and true values are centered around the 45-degree line and the training and testing $R^2$ scores are relatively close to each other, indicating that the base models are neither underfitting nor overfitting. Compared to RF, GBDT and XGBoost have a better fit over the observed data. XGBoost gives the best and generalized prediction when EUR is lower than $3.0 \times 10^8 m^3$, however it can hardly handle larger EUR conditions. Among the three models, GBDT is of the best training accuracy but the poorest testing performance. These base models are of various strengths and weaknesses in different aspects.

A meta model is then constructed over the base models. As shown in Fig. 9d, the stacked model is of higher testing accuracy and keeps a good balance between training and testing to avoid overfitting. In addition to the $R^2$ score, mean squared error (MSE) and mean absolute error (MAE) metrics are also calculated for comparison. To alleviate the effects of randomness, the results are obtained from 100 stochastic experiments. As shown in Table 3, the metrics are consistent in model performance evaluation. The four models perform similarly in the training process. In the testing process, the stacked model obviously outperforms other base models. Not only are the mean metrics more in line with expectation, but the standard deviations are also lower, indicating that the stacked model is more stable. Trade-offs are made among the base models and their advantages are integrated in the stacked model. Therefore, the improved stacked model is used for EUR prediction in the following analysis which is more accurate and robust.

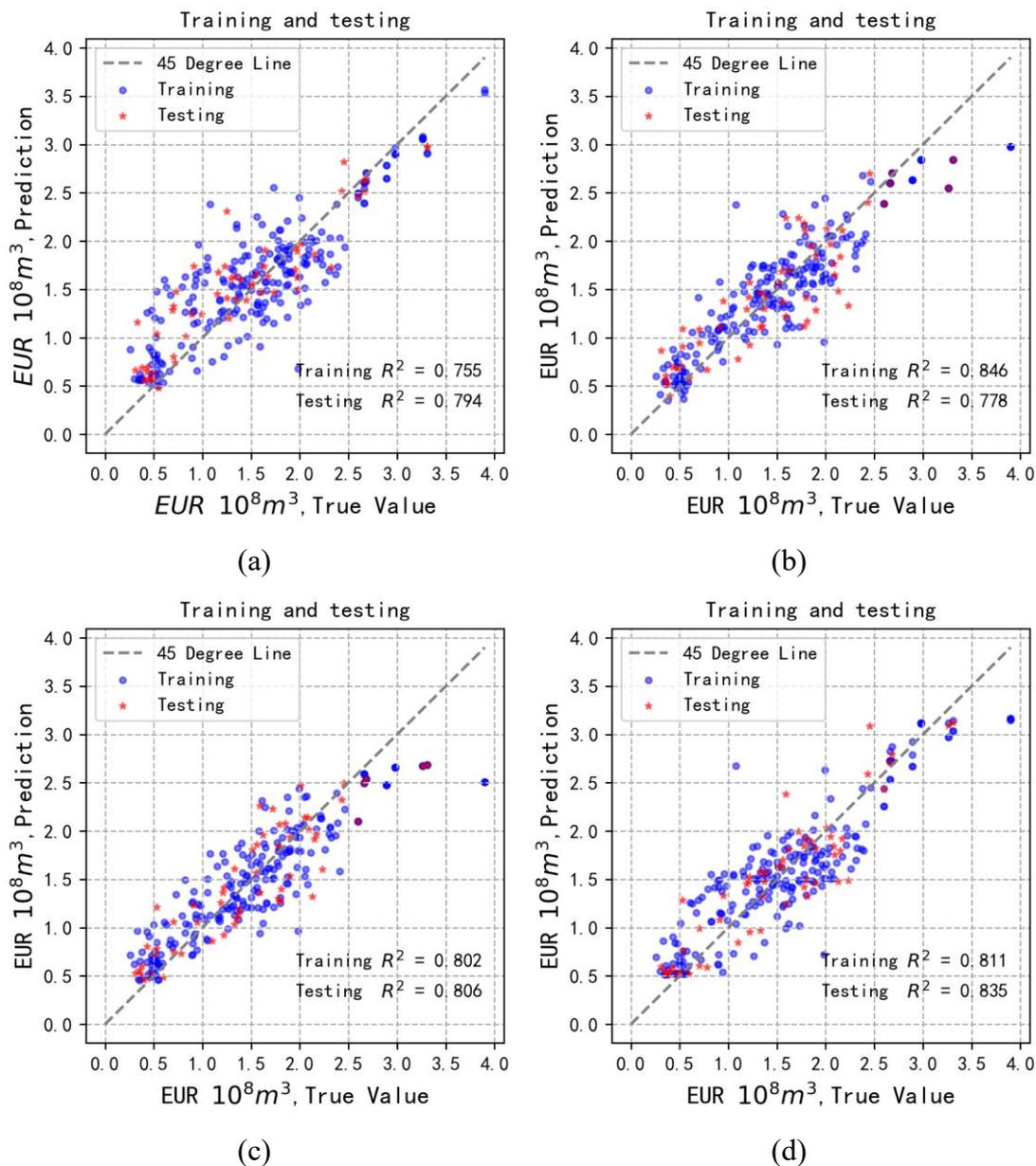

**Fig. 9.** EUR predictions obtained from (a) RF, (b) GBDT, (c) XGBoost and (d) the stacked model versus true values.

**Table 3.** Metrics of RF, GBDT, XGBoost and the stacked model. The value represents the mean±standard deviation of the result obtained from 100 stochastic experiments.

|  | $R^2$ score | | MSE | | MAE | |
| --- | --- | --- | --- | --- | --- | --- |
|  | train | test | train | test | train | test |
| RF | 0.747±0.011 | 0.652±0.050 | 0.160±0.007 | 0.218±0.044 | 0.317±0.008 | 0.371±0.032 |
| GBDT | 0.858±0.015 | 0.703±0.051 | 0.090±0.008 | 0.187±0.041 | 0.221±0.008 | 0.332±0.034 |
| XGBoost | 0.810±0.012 | 0.682±0.064 | 0.120±0.006 | 0.202±0.061 | 0.256±0.008 | 0.342±0.040 |
| Stacked | 0.791±0.027 | 0.743±0.051 | 0.132±0.015 | 0.161±0.038 | 0.276±0.014 | 0.308±0.032 |

With the EUR prediction model, sweet spots which are mutually governed by multiple geological and engineering factors can be better investigated through performance-diagnostic and visualization techniques. At first, single factor analysis is carried out with the aid of the ICE approach introduced in section 2.3. Three dominant factors are studied and shown in Fig. 10, where the target factor changes within the range of research while others remain constant. The sweet spot for any single factor can be directly observed from the ICE curves. In Fig. 10a, EUR increases about 20% when pressure coefficient is larger than 1.5, indicating that formation pressure larger than normal hydrostatic pressure is favorable for resource recovery in the study area. 10%~20% hydrocarbon saturation and 4.5%~5.5% porosity are two crucial intervals where the average EUR changes obviously. A plateau is always reached because there must be an upper limit for the influence on EUR.

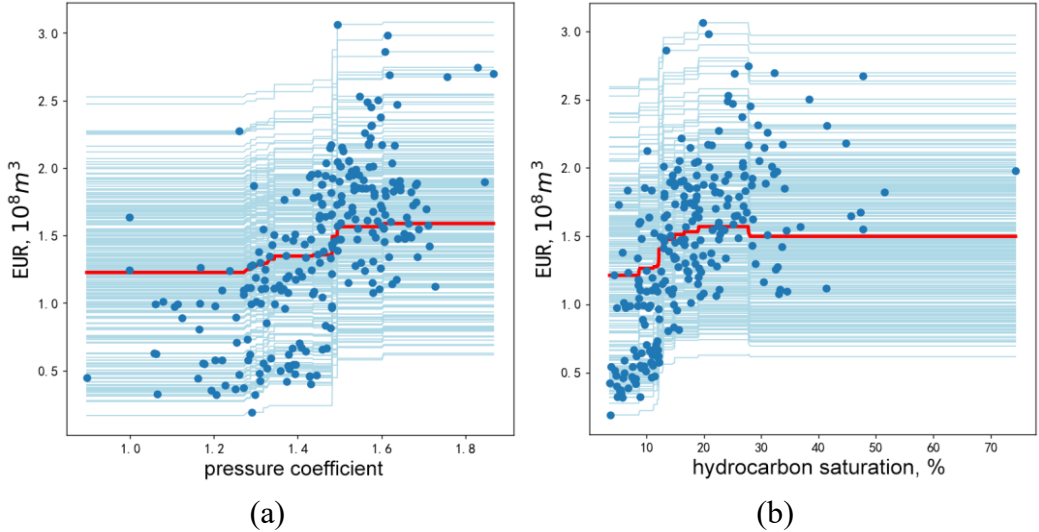

(a)          (b)

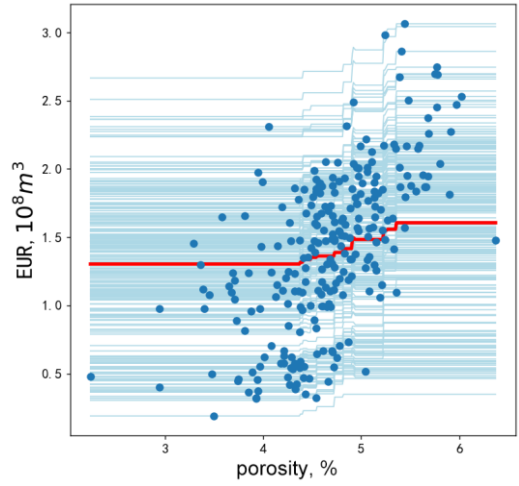

(c)

**Fig. 10.** ICE curves of single factor analysis for (a) formation pressure coefficient, (b) hydrocarbon saturation and (c) porosity. Each dot represents a sample from training dataset and the curve passing through depicts the predicted EUR trend with the target factor changing. The red curve represents the domain average.

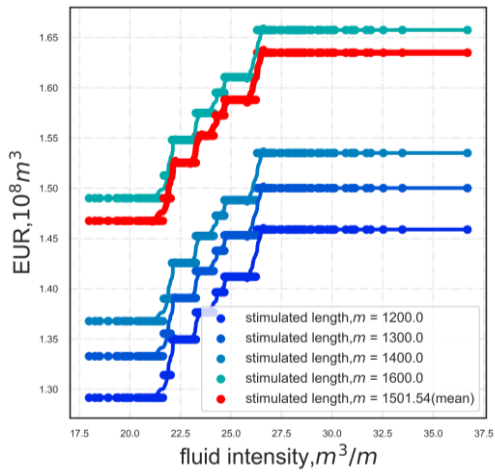
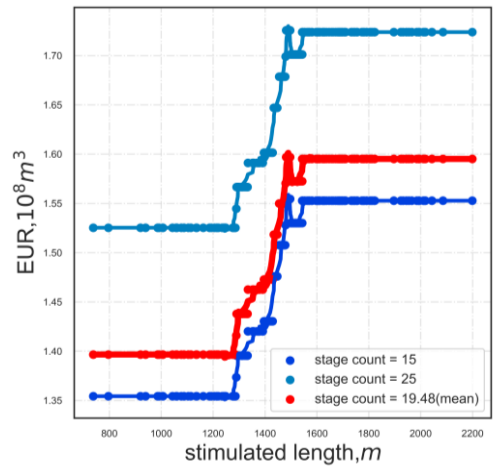

(a)                         (b)

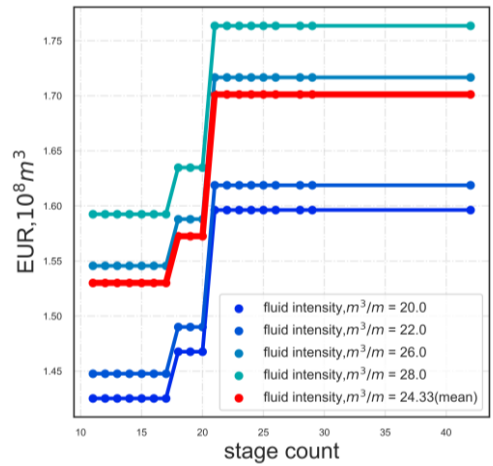

(c)

**Fig. 11.** Projected curves of ICE surface for double factor analysis. Pairwise factor combinations among fluid intensity, stimulated length and stage count are studied.

When designing the development plan, it's impossible to change one factor without altering another. For example, with stimulated length increasing in well planning, stage count should increase as well to facilitate fracturing operation. Thus we further implement multi factor analysis and start from double factors. In the ICE visualization, two factors and the corresponding predictions form a surface in the spanned space. Fig. 11a shows the joint effect of fluid intensity and stimulated length on EUR. The ICE surface is projected to multiple curves for the convenience of observation. The EUR gradually improves by cooperatively increasing these two factors and a bottleneck is reached when stimulated length is longer than 1,500$m$ and fluid intensity is larger than 27$m^3/m$. It reminds us that there is little benefit to continuously strengthen these two factors. Attention should be paid to other potential factors such as stage count (Fig. 11b and 11c). In such situation, double factor analysis won't meet the demand, thus we need to perform triple factor analysis.

An ICE volume is formed by the triple factors and the corresponding EUR predictions, which can be projected to surfaces for illustration. As shown in Fig. 12, best EUR is reached when stage count is 25 with fluid intensity and stimulated length in their favorable zone. However, it's probably not the best choice in consideration of cost, benefit and efficiency. Modifications are recommended to be made in the area where the slope is large, since a small investment may bring significant production elevation. With the aid of the performance diagnosis, one can set proper expectations for development plan implementation and allocate resources heuristically. In conclusion, the visualization techniques provide more insights to sweet spot identification and development evaluation.

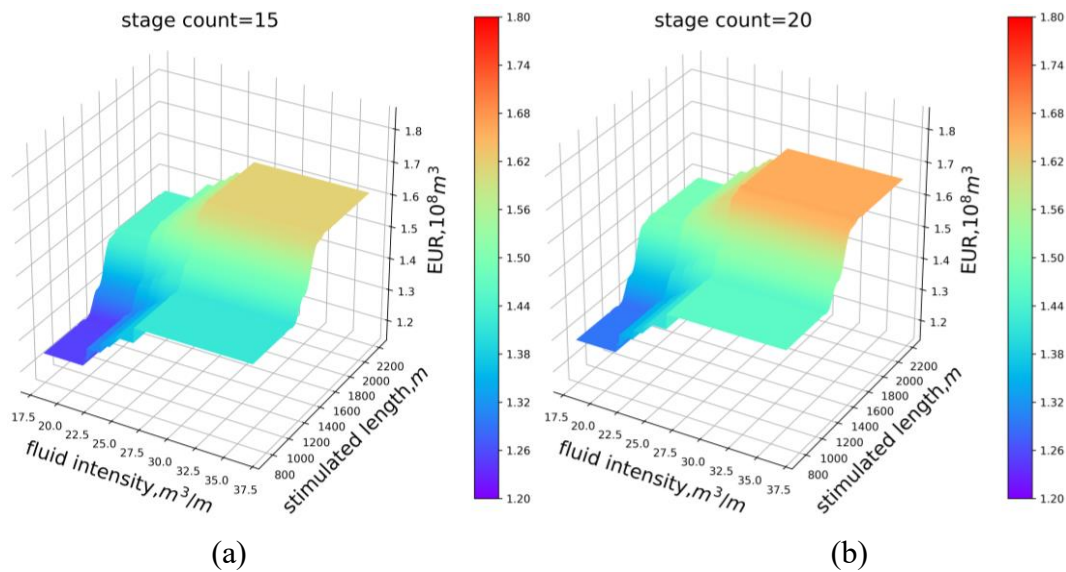

(a)  (b)

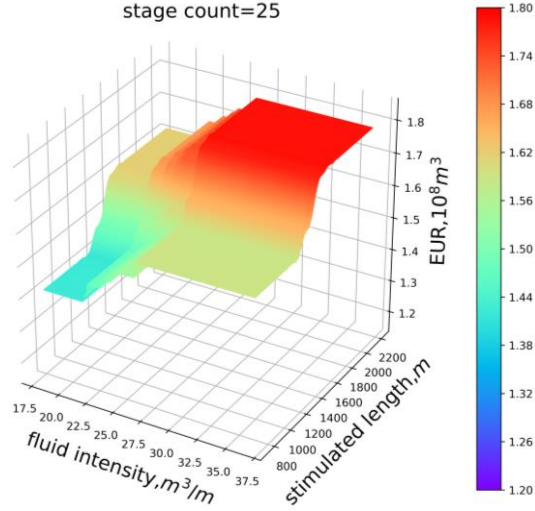

(c)

**Fig. 12.** Projected surfaces of ICE volume for triple factor analysis. Fluid intensity and stimulated length change with stage count being (a) 15, (b) 20 and (c) 25.

Nevertheless, when it comes to development plan optimization, the performance-diagnostic technique is very much restricted, especially when more factors are involved simultaneously. On the one hand, the cost of exhaustive search in the high dimensional space is very expensive. On the other hand, the results can hardly be visualized in an intuitive and comprehensive way. Therefore proper optimization algorithms need to be applied to address this problem.

**3.4. Development plan optimization**

In this section, we implement and test the performance of the three optimization algorithms mentioned in section 2.4. Throughout our study, the same set of hyperparameters for each optimization algorithm are used. For PSO, swarm size $P$ is set as 10, $\omega$, $c_1$, and $c_2$ are 0.729, 1.494, and 1.494 respectively. These values demonstrate good performance in a wide range of problems. For DE, the population size $P$ is also set as 10, and the crossover constant is 0.7. For BO, a Gaussian process is used to construct the surrogate and expected improvement is used as the acquisition function. Meta optimization is omitted since the problem is relatively simple.

We apply PSO, DE, and BO to optimize five engineering parameters of the hydraulic fracturing design: stimulated length, stage count, fracturing fluid intensity, proppant intensity, and angle to the minimum horizontal stress (angle to Hmin). The optimization problem aims at maximizing the EUR of each single well. For illustration, how EUR evolves with model inference times growing for a single well is shown in Fig. 13. The three algorithms converge rapidly and reach similar optimization targets.

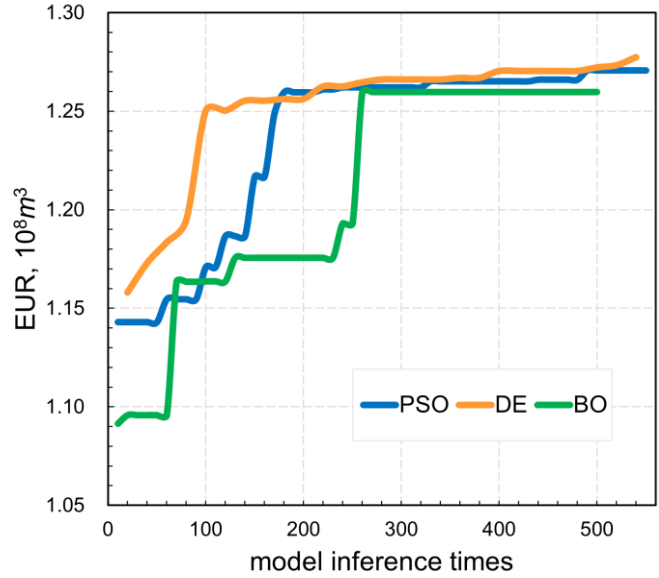

**Fig 13.** Evolution of EUR with model inference times growing for a single well using three derivative-free optimization algorithms.

For comparison, we test the optimization performance with low-EUR wells from both the main and non-main sectors in the Jiaoshiba area. The EURs are lower in the non-main sector due to the inferior geologic conditions. The optimization results for main and non-main sectors are summarized and compared to the original designs in Table 4 and 5 respectively. The fracturing operation is intensified but the engineering parameters are adjusted in different manners. Improvement strategies for both sectors are visualized and compared in Fig. 14. For the main sector, angle to Hmin and stage count are increased the most, and the result suggests that the average EUR could potentially increase by 49.2%. Noting that even though angle to Hmin is increased by 11.2 times, the absolute increasement is relatively small. For the non-main sector, angle to Hmin is decreased, and proppant intensity are increased the most, where EUR could increase by 39.3%. The three algorithms always reach similar optimization targets, but the fracturing designs are slightly different due to the local maxima, fluctuation and nonconvexity of the searching space.

**Table 4.** Comparison of the average values of original hydraulic fracturing parameters and the optimized parameters in the main sector.

|  | stimulated length, $m$ | stage count | fluid intensity, $m^3/m$ | proppant intensity, $m^3/m$ | angle to Hmin, | EUR, $10^8 m^3$ |
|---|---|---|---|---|---|---|
| Original design | 1444.2 | 19 | 26.1 | 0.62 | 0.5 | 0.86 |
| BO | 1913.0 | 28 | 29.4 | 1.09 | 8.5 | 1.24 |
| PSO | 1926.8 | 29 | 29.0 | 0.83 | 1.6 | 1.30 |
| DE | 1766.6 | 28 | 30.4 | 0.76 | 6.7 | 1.31 |
| Average change | +29.4% | +49.1% | +13.4% | +44.1% | *11.2 | +49.2% |

**Table 5**. Comparison of the average values of original hydraulic fracturing parameters and the optimized parameters in the non-main sector.

|  | stimulated length, $m$ | stage count | fluid intensity, $m^3/m$ | proppant intensity, $m^3/m$ | angle to Hmin, | EUR, $10^8 m^3$ |
|---|---|---|---|---|---|---|
| Original design | 1364.4 | 17 | 23.4 | 0.58 | 15.0 | 0.34 |
| BO | 1948.6 | 30 | 32.6 | 1.37 | 2.6 | 0.47 |
| PSO | 1914.5 | 30 | 32.3 | 1.12 | 1.0 | 0.48 |
| DE | 1899.4 | 27 | 30.5 | 0.98 | 4.7 | 0.48 |
| Average change | +40.8% | +70.6% | +52.5% | +99.4% | -81.6% | +39.3% |

The optimal set of engineering parameters can be effectively searched out of a large number of possible combinations with derivative-free optimization algorithms. Customized optimization suggestions are proposed for each single well in consideration of the specific geologic conditions and the interactions among different engineering parameters. Although the algorithms require repetitive computations of the objective function, the optimization process is highly efficiently with credit to the machine learning-based model, which provides instant inference for EUR prediction.

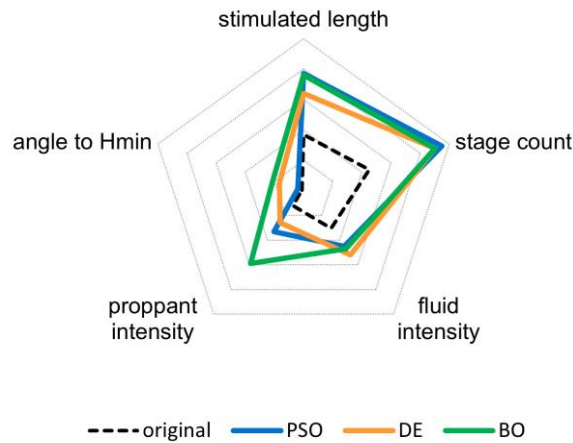

(a)

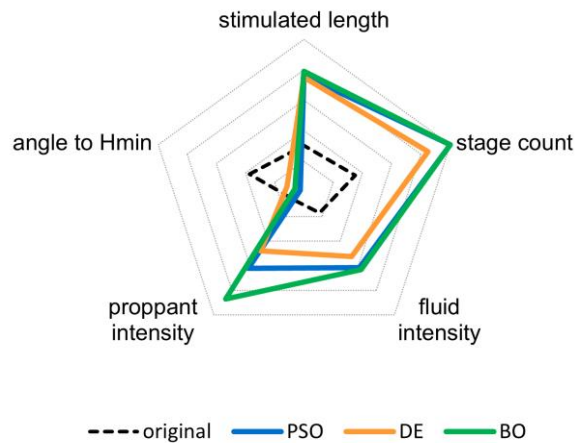

(b)

**Fig 14.** Radar charts of the original and optimized engineering parameters for low-EUR wells from (a) main and (b) non-main sectors using three optimization algorithms.

## 4. Conclusions

In this research, a novel hybrid data-driven framework is proposed for shale gas production performance analysis, prediction and optimization. At first, dominant geological and engineering factors are analyzed and quantified by game theory method. Then a stacked model embodying multi machine learning models are trained for production prediction. At last, suggestions for production enhancement are proposed by optimizing the trained model. The complete workflow is validated with actual field data from Fuling shale gas field. According to the analysis results, the following conclusions are drawn:

(1) Dominant factors can be quantitatively evaluated and ranked by SHAP analysis in the proposed framework. Compared to traditional methods, the proposed method is much more informative and interpretable. Factor impacts on production can be clearly depicted for not only the whole region but also each production well, with which one can gain more enlightenments to shale gas exploration and development.

(2) The stacked model is capable of making accurate predictions for production wells in various geological and engineering conditions. On its basis, the changing patterns can be further diagnosed and visualized by ICE method. One can directly identify potential sweet spots and favorable engineering designs from the ICE curves, surfaces and volumes. The intuitive visualizations can provide instrumental reference for resource evaluation and recovery.

(3) Optimal engineering parameters can be efficiently obtained by derivative-free optimization algorithms. The parameters are optimized cooperatively to find the global optimal solution. The method provides customized suggestions for each well to realize their maximal potential. One can acquire useful insights for estimating potential production and improving the practice in developing plan design.

Compare to traditional analysis approaches which are ambiguous and obscure, the proposed analysis framework is comprehensive and quantitative, from which much more informative and interpretable conclusions can be drawn. For example, in the Fuling shale gas case, we can conclude from the analysis results that geological factors such as formation depth, hydrocarbon saturation, and pressure have greater impacts on the EUR. Stimulated length is the most significant engineering factor. The quantitative results reveal that the effects of most factors always reach a plateau or a peak at some favorable intervals, rather than monotonically increasing or decreasing as expected in traditional qualitative analysis. Note that the data-driven framework is not restricted to production performance analysis for shale gas. Through some modifications, the method can be further applied for cost and profit accounting, investment decision, etc. for different type of resources. Further research will be carried out in our future works.

**Appendix A: Example of Shapley value calculation**

A simple card game example is shown in Fig. A1 to exhibit the standard calculation process of Shapley value (Michael, 2020). Three players cooperative with each other to make a total payoff. How to allocate the money fairly according to the practical contribution of each player is the key problem. If no player is involved, no payoff is obtained $f_\varnothing = \$0$. When each player plays individually, their contributions are given as $f_A = \$7$, $f_B = \$4$, and $f_C = \$6$. When two players play together, we have $f_{AB} = \$7$, $f_{AC} = \$15$, and $f_{BC} = \$9$. When three players are all participated, a total payoff $f_{ABC} = \$19$ can be made. According to Eq. 1.1, the Shapley value for player A is:

$$\begin{aligned}\phi_A = &\ \frac{0!(3-0-1)!}{3!}(f_A - f_\varnothing) &\Leftarrow S = \{\varnothing\} \\ &+ \frac{1!(3-1-1)!}{3!}(f_{AB} - f_B) &\Leftarrow S = \{B\} \\ &+ \frac{1!(3-1-1)!}{3!}(f_{AC} - f_C) &\Leftarrow S = \{C\} \\ &+ \frac{2!(3-2-1)!}{3!}(f_{ABC} - f_{BC}) &\Leftarrow S = \{B,C\} \\ =&\ \$7.7, \end{aligned} \quad (1.14)$$

which represents the marginal contribution of player A. Similarly, Shapley value for player B and C can be obtained: $\phi_B = \$3.2$, $\phi_C = \$8.2$. According to the additivity property illustrated in Eq. 1.2, we have:

$$\begin{aligned} g(z') &= \phi_0 + \sum_{i=\{A,B,C\}} \phi_i z'_i \\ &= f_\varnothing + \phi_A + \phi_B + \phi_C \\ &= \$0 + \$7.7 + \$3.2 + \$8.2 \\ &= f_{ABC} \end{aligned} \quad (1.15)$$

where $z' = \{1,1,1\}$ representing the three players are all involved. Then the total payoff should be fairly assigned to each player according to their marginal contributions quantified by Shapley values. The simple case validates that Shapley value is a powerful attribution for revenue allocation and model explanation.

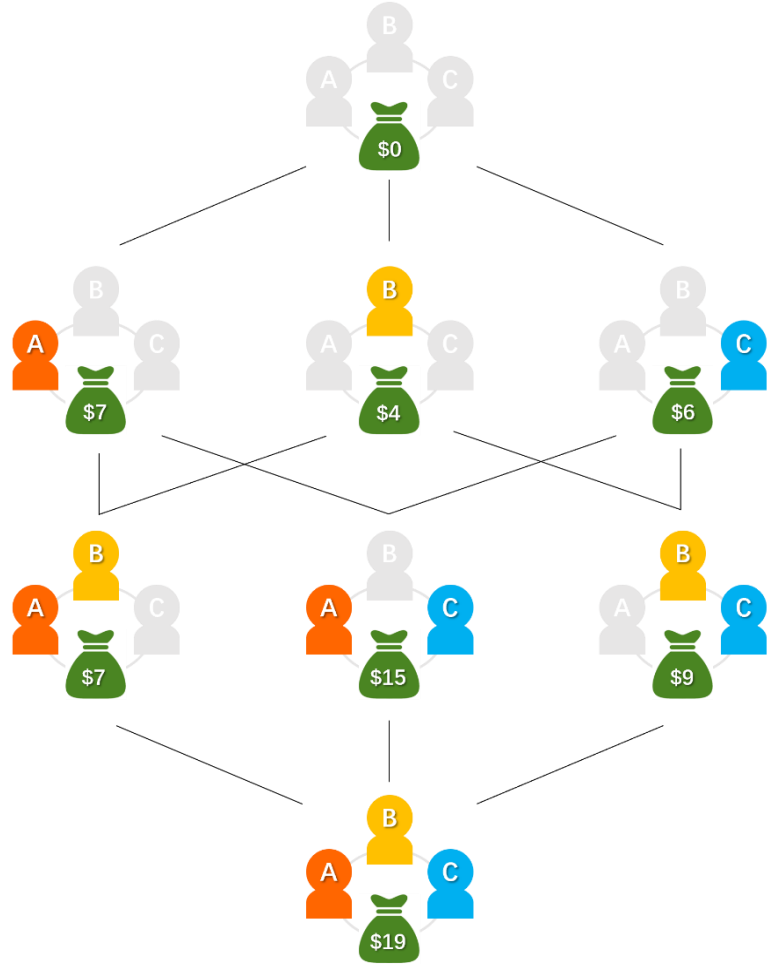

**Fig A1.** Example of a simple three-player coalitional card game.

However, in practical applications, the conditional payoff (e.g., $f_A, f_{BC}$) is rarely explicitly given but requires repetitive model running. In theory, $2^M$ model differences have to be computed where $M$ is the number of players or input features. For example, in this simple three-player card game, $2^3=8$ conditional payoff states are required for Shapley value calculation. When there are more players or features involved, the exponentially increased computational efforts will be unaffordable. Thus, the SHapley Additive exPlanation (SHAP) value is a necessary alternative and approximation to Shapley value, which can be efficiently computed without running models exhaustively. There are several model type specific SHAP approximations for different situation, including kernel SHAP, linear SHAP, low order SHAP, max SHAP, deep SHAP, and tree SHAP (Lundberg et al., 2020; Lundberg et al., 2018; Lundberg and Lee, 2017). It can be proved that the above methods are the only possible consistent, locally accurate methods that obey the missingness property and uses conditional dependence to measure missingness (Lundberg et al., 2018).

**Appendix B: Summary of raw data collected from the first period production project of Jiaoshiba area, Fuling shale gas field**

| Category | Name | Unit | Optimizable | Data amount |
|---|---|---|---|---|
| Geologic | Formation depth | m | N | 254 |
| | TOC | % | N | 254 |
| | Gas content | $m^3/ton$ | N | 248 |
| | Porosity | % | N | 254 |
| | Permeability | mD | N | 254 |
| | Hydrocarbon saturation | % | N | 253 |
| | Methane saturation | % | N | 253 |
| | Gas saturation | % | N | 252 |
| | Quartz content | % | N | 190 |
| | Formation pressure | MPa | N | 254 |
| | Tectonic curvature | - | N | 254 |
| | Initial pressure | MPa | N | 170 |
| | Breakdown pressure | MPa | N | 254 |
| | Sector | - | - | 254 |
| Drilling | Well name | - | - | 254 |
| | Latitude | - | - | 254 |
| | Longitude | - | - | 254 |
| | TVD | m | N | 254 |
| | Kelly bushing | m | N | 254 |
| | Lateral length | m | Y | 254 |
| | Layer 1 penetration | % | Y | 214 |
| | Layer 2 penetration | % | Y | 221 |
| | Layer 3 penetration | % | Y | 246 |
| | Elevation change between | m | Y | 254 |
| | Angel to min. horizontal stress | | Y | 254 |
| | ROP | min/m | Y | 253 |
| Completion | Stimulated length | m | Y | 254 |
| | Stage count | - | Y | 254 |
| | Perforation count | - | Y | 254 |
| | Slick water volume | $m^3$ | Y | 254 |
| | Gel volume | $m^3$ | Y | 254 |
| | Fracturing fluid volume | $m^3$ | Y | 254 |
| | 100 mesh volume | $m^3$ | Y | 254 |
| | 40/70 mesh volume | $m^3$ | Y | 254 |
| | 30/50 volume | $m^3$ | Y | 244 |
| | Proppant volume | $m^3$ | Y | 254 |
| | ISIP | MPa | N | 254 |
| Production | Peak test production | $10^4 m^3$ | - | 254 |
| performance | Absolute open flow | $10^4 m^3$ | - | 254 |
| | Casing pressure | $MPa$ | - | 254 |
| | 3-month cumulative | $10^4 m^3$ | - | 252 |
| | 6-month cumulative | $10^4 m^3$ | - | 251 |
| | 12-month cumulative | $10^4 m^3$ | - | 249 |
| | EUR | $10^8 m^3$ | - | 254 |